\documentclass[sigconf]{acmart}

\usepackage{geometry}
\usepackage{listings}
\usepackage{multicol} 
\usepackage{float}
\usepackage{hyperref} 

\AtBeginDocument{%
  \providecommand\BibTeX{{%
    \normalfont B\kern-0.5em{\scshape i\kern-0.25em b}\kern-0.8em\TeX}}}

\copyrightyear{2022}
\acmYear{2022}
\acmBooktitle{The 1st ACM SIGSPATIAL International Workshop on Geospatial Knowledge Graphs (GeoKG '22), November 1, 2022, Seattle, WA, USA}

\setcopyright{none} 

\begin{document}



\title[Measuring Network Resilience via Geospatial Knowledge Graph]{Measuring Network Resilience via Geospatial Knowledge Graph: a Case Study of the US Multi-Commodity Flow Network}

\author{Jinmeng Rao}
\affiliation{%
  \institution{Geospatial Data Science Lab, University of Wisconsin-Madison}
  \city{Madison}
  \country{USA}}
\email{jinmeng.rao@wisc.edu}

\author{Song Gao}
\affiliation{%
  \institution{Geospatial Data Science Lab, University of Wisconsin-Madison}
  \city{Madison}
  \country{USA}
}
\email{song.gao@wisc.edu}

\author{Michelle Miller}
\affiliation{%
  \institution{Center for Integrated Agricultural Systems, University of Wisconsin-Madison}
  \city{Madison}
  \country{USA}
}
\email{mmmille6@wisc.edu}

\author{Alfonso Morales}
\affiliation{%
  \institution{Department of Planning and Landscape Architecture, University of Wisconsin-Madison}
  \city{Madison}
  \country{USA}
}
\email{morales1@wisc.edu}

\renewcommand{\shortauthors}{Rao et al.}


\begin{abstract}
Quantifying the resilience in the food system is important for food security issues. In this work, we present a geospatial knowledge graph (GeoKG)-based method for measuring the resilience of a multi-commodity flow network. Specifically, we develop a CFS-GeoKG ontology to describe geospatial semantics of a multi-commodity flow network comprehensively, and design resilience metrics that measure the node-level and network-level dependence of single-sourcing, distant, or non-adjacent suppliers/customers in food supply chains. We conduct a case study of the US state-level agricultural multi-commodity flow network with hierarchical commodity types. The results indicate that, by leveraging GeoKG, our method supports measuring both node-level and network-level resilience across space and over time and also helps discover concentration patterns of agricultural resources in the spatial network at different geographic scales.
\end{abstract}





\maketitle

\section{Introduction}

According to the State of Food Security and Nutrition in the World report in 2022 by the Food and Agriculture Organization of the United Nations \cite{FAO2022}, the number of people affected by global hunger has increased by 150 million since the COVID-19 outbreak, reaching 828 million in 2021. 11.7\% of the global population faced food insecurity nowadays. Resilient and sustainable food supply chain networks may benefit the production, delivery, and consumption of agricultural products, thereby showing promises in mitigating global hunger and critical for regional and global food security \citep{suweis2015resilience,miller2021identifying,karakoc2021complex}. However, food supply chain networks involve multidimensional interactions and complex decisions.  All participants in supply chains contend with diverse food demands, transportation time and cost, sales strategy, natural conditions, etc. This means that understanding supply chain resilience is a grand challenge \citep{chaturvedi2014securing, yadav2022systematic}.


Recently, different measures have been proposed in complex (spatial) networks \citep{gao2016universal,dey2019network,li2016resilience} and applied in understanding the resilience and the resilience-efficiency trade-off in the global food trade network \citep{fair2017dynamics,puma2019resilience,tu2019impact,karakoc2021complex}. For example, \citet{fair2017dynamics} built a dynamic model of the global wheat trade network and explored the resilience of the network to targeted attacks on edges. \citet{tu2019impact} measured the connectivity, structure, and modularity of a global food trade network and investigated their association with the resilience of the network. \citet{karakoc2021complex} studied topological and weighted resilience and efficiency metrics and proposed a complex network framework for evaluating the trade-off between efficiency and resilience of food trade networks. However, it is still hard to comprehensively assess the food system resilience with complex multidimensional information (e.g., multiple commodity types, suppliers and customers, geographic proximity, and at different scales). With the recent advances in geospatial knowledge graphs (GeoKG) \citep{janowicz2022know,mai2020se}, the importance of spatial concepts (e.g., the scale of geographic entities and spatial dependence) has been addressed in knowledge discovery, semantic reasoning, etc. Such concepts can be integrated with food systems \citep{hollander2020toward} to help decision makers understand and improve the structure and resilience of food supply chain networks, thereby safeguarding local, regional, and global food security.

In this work, we present a novel method for measuring network resilience via GeoKG and apply it for understanding the node-level and network-level resilience of the US agricultural multi-commodity flow network at different geographic scales. By leveraging GeoKG, we are able to integrate several factors into resilience metrics (i.e., commodity value, supplier or customer commodity diversity, average transport miles, and geographic adjacency) and comprehensively measure the single-sourcing dependence of suppliers, customers, and commodity types as well as the geographic proximity and adjacency (spatial dependence) in the network. Such comprehensive resilience metrics may not only allow a good assessment of the resilience of the entire network or the nodes themselves when facing disruption risks, but also help us discover trends in the dispersion or concentration of agricultural resources at different geographic scales.

The remainder of the paper is organized as follows: In Section 2, we first introduce the ontology design of CFS-GeoKG, a geographic knowledge graph for storing and querying geospatial semantics of multi-commodity flow data. We then propose two resilience metrics for the comprehensive measurement of node-level and network-level resilience via CFS-GeoKG which considers commodity value and diversity, average transport miles, and geographic adjacency. In Section 3, we report and discuss the results from a case study of the US agricultural multi-commodity flow network using our proposed methods. In section 4, we discuss the limits associated with our data source and metric design, and we summarize corresponding future research directions. Finally, in Section 5, we conclude this work and outline future works.

\section{Methods}

In this section, we first present a GeoKG ontology design: CFS-GeoKG, which describes the US multi-commodity flow network based on the US Commodity Flow Survey (CFS) data\footnote{\url{https://www.census.gov/programs-surveys/cfs.html}}. Then we introduce several network resilience evaluation metrics based on the designed CFS-GeoKG, which comprehensively consider multi-dimensional semantics such as supplier/customer/commodity diversity, commodity value, geographic adjacency, and average transport mileage in the measurements. We also demonstrate how to calculate the proposed resilience metrics via CFS-GeoKG using the SPARQL query.

\begin{figure*}[h]
	\centering
	\includegraphics[width=0.8\linewidth]{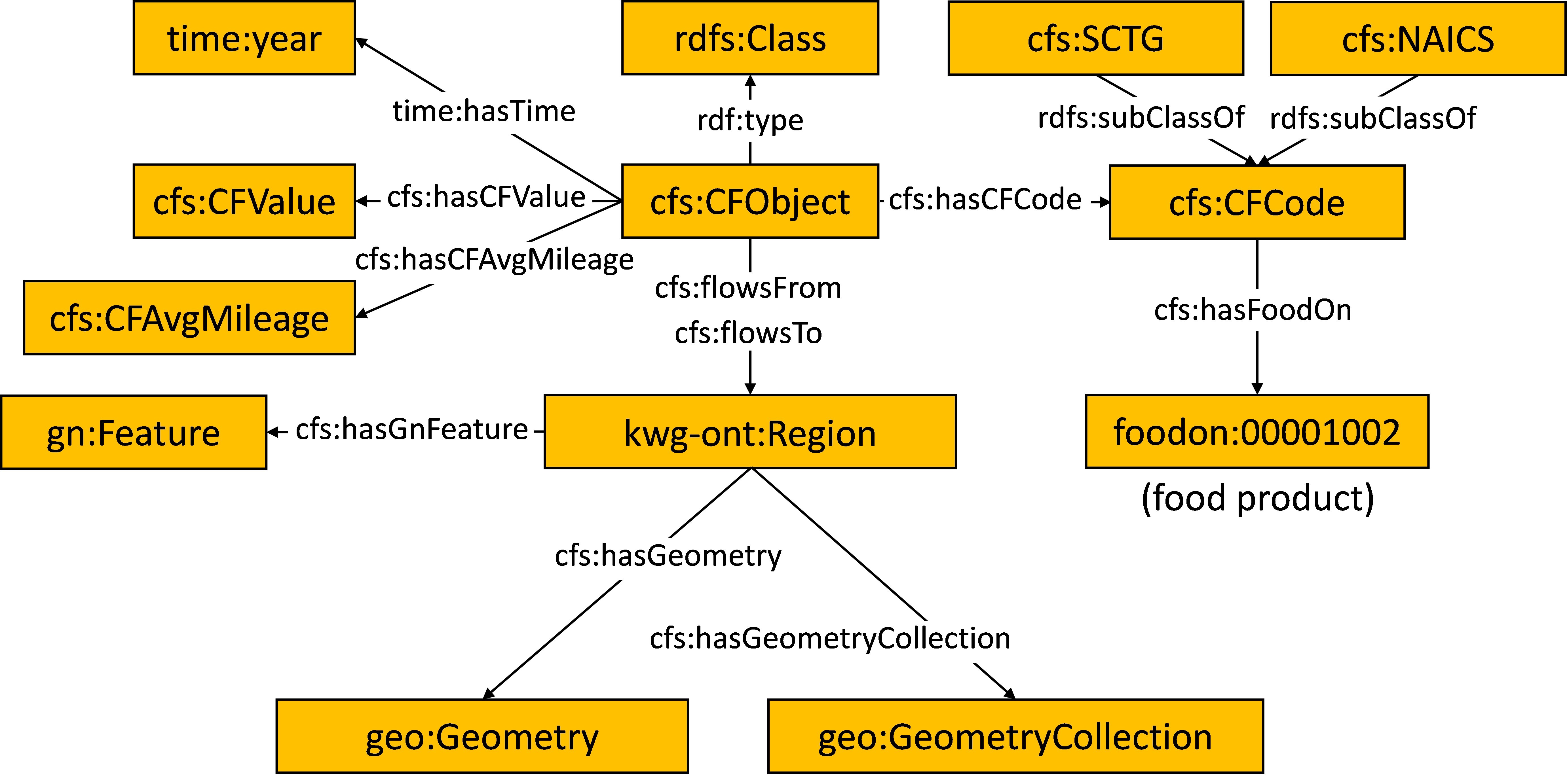}
	\caption{The ontology design of CFS-GeoKG.}
	\label{fig:ontology}
\end{figure*}

\subsection{CFS-GeoKG Ontology}

We design an ontology ``CFS-GeoKG''  to comprehensively describe geospatial semantic information in the US multi-commodity flow network. The CFS-GeoKG ontology is designed to 1) define hierarchical geographical entities and their relations in the multi-commodity flow network; 2) describe hierarchical multi-type commodity flows and their properties; and 3) support network resilience measurement; and 4) link out to and reuse existing semantic ontology design patterns for better interoperability or interlinking existing Linked data and other Web resources \citep{janowicz2014five}. Figure \ref{fig:ontology} presents the ontology design of CFS-GeoKG. Each commodity flow in the network belongs to the class \textit{cfs:CFObject}, which has several essential properties: \textit{cfs:CFValue}, the total value (\$ millions) of the commodity flow; \textit{cfs:AvgMileage}, the average transport mileage (miles) of the commodity flow; \textit{time:year} from the Time Ontology in OWL\footnote{\url{https://www.w3.org/TR/owl-time/}}, the year of the commodity flow; \textit{cfs:CFCode}, the commodity code such as Standard Classification of Transported Goods (SCTG, described by \textit{cfs:SCTG})\footnote{\url{https://bhs.econ.census.gov/bhsphpext/brdsearch/scs_code.html}} or North American Industry Classification System (NAICS, described by \textit{cfs:NAICS})\footnote{\url{https://www.census.gov/naics/}}. Note that the hierarchical structure in commodity types, if exists, can be preserved by adding corresponding data properties to \textit{cfs:CFCode}; and \textit{kwg-ont:Region}\footnote{\url{https://stko-kwg.geog.ucsb.edu/lod/ontology\#Region}} from the KnowWhereGraph ontology \citep{janowicz2022know}, describing hierarchical geographical entities involved in CFS data such as the geographical or administrative origin and destination (e.g., state, region, division, and CFS area) of a commodity flow. Note that the concepts of regions and divisions we employ in the paper are from the Geographic Levels defined by the US Census Bureau\footnote{\url{{https://www.census.gov/programs-surveys/economic-census/guidance-geographies/levels.html}}}.

For each agricultural commodity code (i.e., each food type), if available, we also link it to its corresponding FoodOn ontology class \citep{dooley2018foodon} to add more food information. To support the storage and query of geospatial information, we utilize the GeoSPARQL developed by Open Geospatial Consortium (OGC) that defines a vocabulary for representing geospatial data in resource description framework (RDF) \citep{battle2011geosparql}. Specifically, \textit{kwg-ont:Region} has either a \textit{geo:Geometry} or a \textit{geo:GeometryCollection} depending on its geographical type. For example, a state has one \textit{geo:Geometry} that describes its boundary, area, etc.; a region or division contains multiple states, therefore having a \textit{geo:GeometryCollection} containing multiple \textit{geo:Geometry}. We also utilize the GeoNames ontology \citep{vatant2006geonames} (e.g., assigning a \textit{gn:Feature} for each \textit{kwg-ont:Region}) to support the storage and query of geographical names, geoID, geographical types, etc. Note that we did not show all CFS attributes in Figure \ref{fig:ontology} for simplicity. In fact, more properties such as the total weight of the commodity flow (e.g., \textit{cfs:CFWeight}) can be described in the CFS-GeoKG. With the CFS-GeoKG, we are able to annotate the geographical entities and commodity flow entities as well as their properties and relations in the semantic network, which lays the foundation for a comprehensive network resilience assessments.

\subsection{Network Resilience Metrics}

\begin{figure*}[h]
	\centering
	\includegraphics[width=0.6\linewidth]{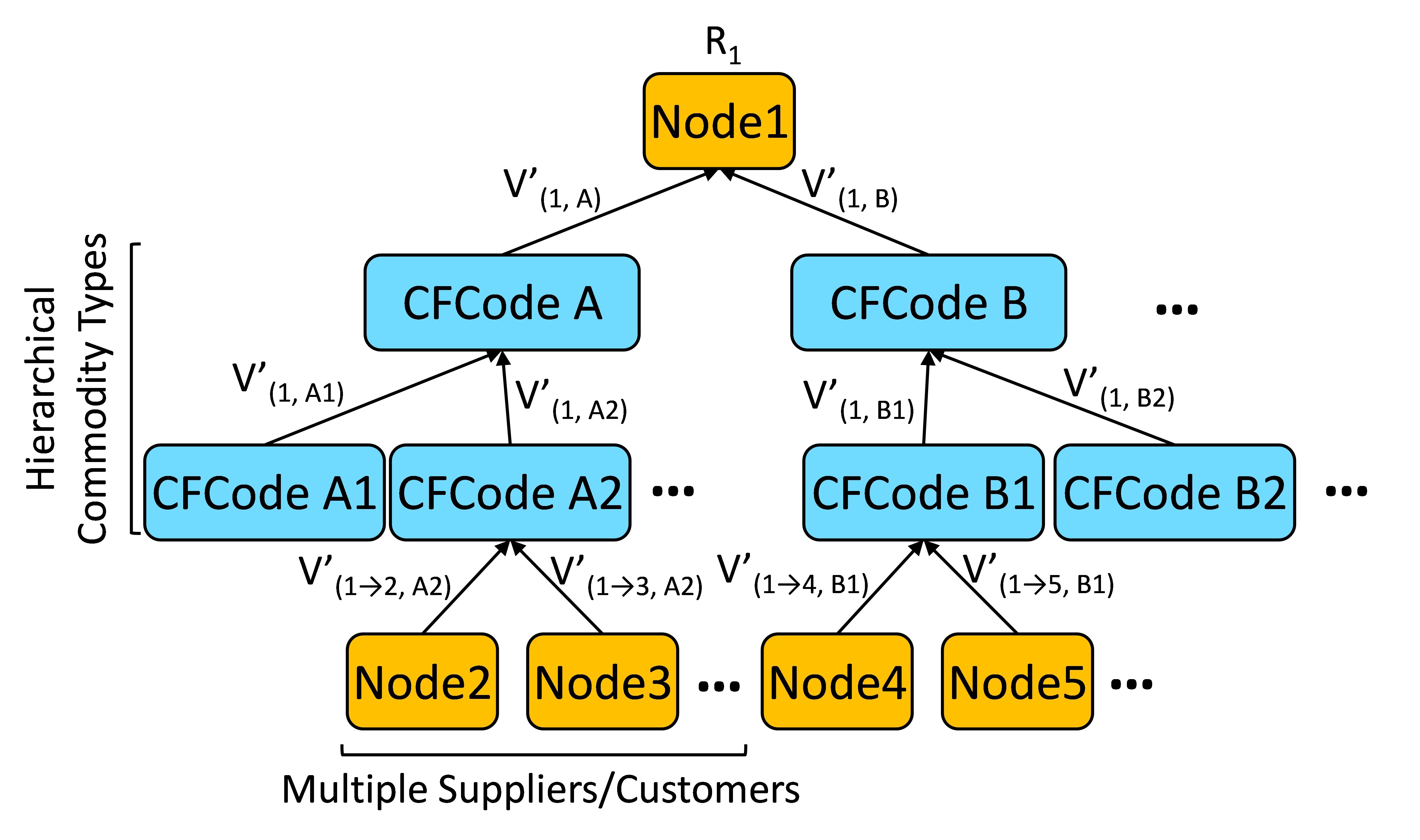}
	\caption{A conceptual graph showing the bottom-up node-level export resilience calculation.}
	\label{fig:node_resilience_calculation}
\end{figure*}

We measure resilience in a multi-commodity flow network mainly based on the extent of dependence on Single-Sourcing Dependence \citep{swift1995preferences, namdar2018supply}. According to different levels of resilience measurement, we propose two types of resilience metrics: \textit{node-level resilience} and \textit{network-level resilience}. Generally speaking, a node (e.g., a state) is less resilient if it tends to rely more on a single supplier, a single customer, or a single commodity type. This is because a centralized risk due to single sourcing leads to a lower fault tolerance. When the largest supplier or customer of the node is removed from the supply chain, or the commodity on which the node is most reliant are out of supply or unsalable, the resulting impact could be devastating for the node's economy (e.g., disruption risks \citep{kamalahmadi2017assessment}). Also, from a geographical perspective, a node is less resilient if it tends to rely more on geographically distant or non-adjacent suppliers or customers. Generally, longer distance transport not only increases transport costs, but also increases transport risks \citep{inman2014product}. We therefore define the resilience of a node as a comprehensive measurement of the dependence on multiple suppliers/customers, commodity types and geographic proximity. In Table \ref{tab:resilience_factors}, we list four factors that affects node-level resilience: Commodity Value (CV), Supplier/Customer/Commodity Diversity (SCCD), Average Transport Mileage (ATM), and Geographic Adjacency (GA). We summarize a bottom-top calculation of node-level resilience in four steps (Figure \ref{fig:node_resilience_calculation}):

\begin{table}[h]
	\centering
	\caption{Factors affecting node resilience.}
	\label{tab:freq}
	\begin{tabular}{p{1cm}p{6.5cm}}
		\toprule
		Factors&Description\\
		\midrule
		CV & The value of a commodity flow\\
		SCCD & The diversity of suppliers/customers/commodities\\
		ATM & The average transport mileage of a commodity flow\\
		GA & The geographic adjacency between the origin and destination of a commodity flow\\
		\bottomrule
	\end{tabular}
	\label{tab:resilience_factors}
\end{table}
\textbf{Combining CV with ATM and GA:} In order to reflect the average
transport mileage and geographic adjacency information in commodity values, we combine them using the following formula:

\begin{equation}
V'_{(i\rightarrow j, c)} = V_{(i\rightarrow j, c)} \times \alpha_{(i\rightarrow j, c)} \times \beta_{(i, j)}
\end{equation}

where $V_{(i\rightarrow j,c)}$ and $\alpha_{(i\rightarrow j,c)}$ denote the value and the factor for average transport mileage of the flow with commodity type $c$ from node $i$ to node $j$, respectively. We set $\alpha_{(i\rightarrow j,c)}=1$ when ATM is shorter than 1 mile, otherwise we set $\alpha_{(i\rightarrow j,c)}$ to be the square root of ATM; $\beta_{(i,j)}$ denotes the factor for geographic adjacency between node $i$ and node $j$. $\beta_{(i,j)} = 0.9$ if node $i$ and node $j$ is geographically adjacent, otherwise $\beta_{(i,j)} = 1$. Note that these factors can be modified based on domain knowledge. Also, in the CFS-GeoKG, CV and ATM can be queried via \textit{cfs:CFValue} and \textit{cfs:AvgMileage}, and GA can be queried using \textit{geo:ehMeet}, the 'Meet' relation in Egenhofer's Topological Relation Family between regions \citep{egenhofer2010family} and implemented by GeoSPARQL.

\textbf{Measuring Single-Supplier/Customer Dependence:} By leveraging Shannon entropy \citep{shannon1948mathematical}, we measure the single-supplier/customer dependence of a node for each commodity type (here we measure the single-customer dependence as an example):

\begin{equation}
H_{(i,c)} = - \sum_{j}p_{j}\log_{2}p_{j}
\end{equation}
\begin{equation}
D_{(i,c)} = 2^{-H_{(i,c)}} = \prod_{j}p_{j}^{p_{j}}
\end{equation}

Where $H_{(i,c)}$ denotes the Shannon entropy of the value distribution of commodity type $c$ centered to node $i$; $p_{j}=V'_{(i\rightarrow j, c)}/\sum_{j}V'_{(i\rightarrow j, c)}$ is the normalized commodity value; $D_{(i,c)}$ denotes the extent of single-customer dependence of node $i$ on commodity type $c$ (i.e., a weighted geometric mean of normalized commodity value distribution \citep{nelson2017assessing}). A larger $D_{(i,c)}$ implies a higher single-customer dependence.

\textbf{Measuring Single-Commodity-Type Dependence:} Building upon the single-customer dependence, we further measure the single-commodity-type dependence of a node. Due to the hierarchy of commodity types (e.g., 'SCTG 02100 Wheat' is a subClass of 'SCTG 02 Cereal Grains') as shown in Figure \ref{fig:node_resilience_calculation}, we first calculate the single-commodity-type dependence for each aggregated commodity type (e.g., CFCode A and B), and then we calculate the overall single-commodity-type dependence of a node:

\begin{equation}
D_{(i,A)} = \prod_{c\in A}p_{(i,c)}^{p_{(i,c)}}
\end{equation}
\begin{equation}
D_{i} = \prod_{A\in Agg}p_{(i,A)}^{p_{(i,A)}}
\end{equation}

where $D_{i}$ is the overall single-commodity-type dependence of node $i$; $Agg$ is a set of aggregated commodity types; $p_{(i,A)}=V'_{(i, A)}/\sum_{A \in Agg}V'_{(i, A)}$ is the normalized value of $V'_{(i, A)}=D_{(i,A)}\sum_{c \in A}V'_{(i, c)}$, which is the value of aggregated commodity $A$ reflecting single-commodity-type dependence; $D_{(i,A)}$ is the single-commodity-type dependence of the aggregated commodity type $A$; $p_{(i,c)}=V'_{(i, c)}/\sum_{c \in A}V'_{(i, c)}$ is the normalized value of $V'_{(i, c)}=D_{(i,c)}\sum_{j}V'_{(i\rightarrow j, c)}$, which is the value of commodity type $c$ reflecting single-customer dependence. Note that in the CFS-GeoKG, aggregated commodity types are stored as data properties of \textit{cfs:CFCode}.

\textbf{Measuring Node-Level Resilience:} Eventually, we get a node-level resilience $R_i$ that comprehensively measures the single-supplier/customer dependence, single-commodity-type dependence, transport distance, and geographic adjacency of the node $i$:

\begin{equation}
R_{i} = 1 - D_{i}\frac{\sum_{A \in Agg}V'_{(i, A)}}{V'_{i}}
\label{eq:R_i}
\end{equation}

where $V'_{i}=\sum_{A \in Agg}\sum_{c \in A}\sum_{j}V'_{(i\rightarrow j, c)}$ denotes the total commodity value of node $i$ combined with average transport miles and geography adjacency information. The higher the $R_i$, the less dependent node $i$ is on single or geographically distant/non-adjacent supplier/customer or single commodity type, thus the higher the resilience. Depending on commodity in-flows and out-flows, we can calculate the node-level import resilience $R^{in}_i$ and export resilience $R^{out}_i$ of node $i$, respectively.

\textbf{Measuring Network-Level Resilience:} It is natural to consider how to extend this node-level resilience metric to a network-level resilience metric. Likewise, network-level resilience also relies on CV, SCCD, ATM, and GA. We refer to the concept of a food trade network resilience proposed by \cite{karakoc2021complex}, where they define network resilience as the resilience to the targeted attack on the major exporter with the most mass supply. We modify this definition as ``the resilience to the targeted attack on the most influential importer/exporter in the multi-commodity network''. Here the most influential importer/exporter may: 1) have a very high total commodity flow value (combined with transport mileage and geographic adjacency) in the multi-commodity network; and 2) have very deep integration with the multi-commodity network (i.e., the business and value are deeply linked to most of the commodity types). The most influential importer/exporter does the best of the two aspects combined, therefore attacking such an importer/exporter will have a huge impact on the vulnerability of entire network. By capturing such a change in the network, we are able to measure the network-level resilience as below:

\begin{equation}
R_{net} = 1 - max(I_{i})
\end{equation}
\begin{equation}
I_{i} = \frac{R_{i}V'_{i}}{\sum_{i}R_{i}V'_{i}}
\end{equation}

where $R_{i}$ denotes the node-level resilience of node $i$, which also implies the extent of integration between node $i$ and the network; $V'_{i}$ denotes the total commodity value of node $i$ described in formula (\ref{eq:R_i}). $I_{i}$ thereby reflects the overall influence of node $i$ on the entire network. Likewise, we can calculate the network-level import resilience $R^{in}_{net}$ and export resilience $R^{out}_{net}$ using commodity in-flows and out-flows, respectively.

\textbf{Resilience Computation via CFS-GeoKG using SPASQL:} The aforementioned resilience metrics can be defined and implemented as on-demand query functions (with the prefix \textit{cfsf:}) \citep{regalia2016volt}, which are used to calculate the node-level and network-level resilience scores via the CFS-GeoKG. Here we give an example of how to retrieve the top-10 states with the highest node-level export resilience in 2017 using the predefined \textit{cfsf:node\_export\_resilience} query function. This function takes a \textit{kwg-ont:Region} node and three other parameters (i.e., year, ATM and GA factors) as inputs, and return the export resilience of that node as the output:
\vspace{1em}
\begin{lstlisting}[
tabsize=1, 
showspaces=false, 
showstringspaces=false,
basicstyle=\footnotesize, 
frame=tbrl, %t: top, r, b, l
numbers=left, 
numberstyle=\tiny, 
numberblanklines=false,
xleftmargin=0.25cm,
xrightmargin=0.25cm,
columns=flexible
]
cfsf:node_export_resilience (node: kwg-ont:Region,
year: time:year,
atm: xsd:string,
ga: xsd:double): xsd:double
\end{lstlisting}
\vspace{1em}
The example SPARQL query is as below:
\vspace{1em}
\begin{lstlisting}[
language=SQL,
tabsize=1, 
showspaces=false,
keywordstyle=\ttfamily\bfseries,
showstringspaces=false,
basicstyle=\footnotesize, 
frame=tbrl, %t: top, r, b, l
numbers=left, 
numberstyle=\tiny, 
numberblanklines=false,
xleftmargin=0.25cm,
xrightmargin=0.25cm,
columns=flexible
]
SELECT ?state (cfsf:node_export_resilience(?state, 2017, 'sqrt', 0.9)
AS ?expResilience)
WHERE {
?state rdf:type kwg-ont:Region.
?state cfs:hasGnFeature ?gnFeature.
?gnFeature gn:featureCode 'ADM1'.
}
ORDER BY DESC(?expResilience)
LIMIT 10
\end{lstlisting}
\vspace{1em}

where \textit{?state} denotes the nodes in the CFS-GeoKG we select. We first select the state-level \textit{kwg-ont:Region} nodes with \textit{gn:featureCode} as 'ADM1', and then we calculate their node-level export resilience via the \textit{cfsf:node\_export\_resilience} query function, where we set \textit{year} as 2017 for calculating resilience scores in 2017, \textit{atm} as 'sqrt' to use the square root of ATM for the $\alpha$ factor reflecting average transport mileage, and \textit{ga} as 0.9 for the $\beta$ factor reflecting geographic adjacency. Lastly, we sort the results based on the calculated export resilience in descending order and return the first 10 results. Similarly, the node-level import resilience and the network-level resilience measures can also be calculated on-the-fly with predefined SPARQL query functions. The key consideration of on-demand computation vs. pre-compute  is because we can calibrate the parameters and update resilience scores to see the changes, especially when we want to investigate how a change in node property would affect the network resilience.

\section{Results}

\begin{figure*}[h]
	\centering
	\includegraphics[width=0.9\linewidth]{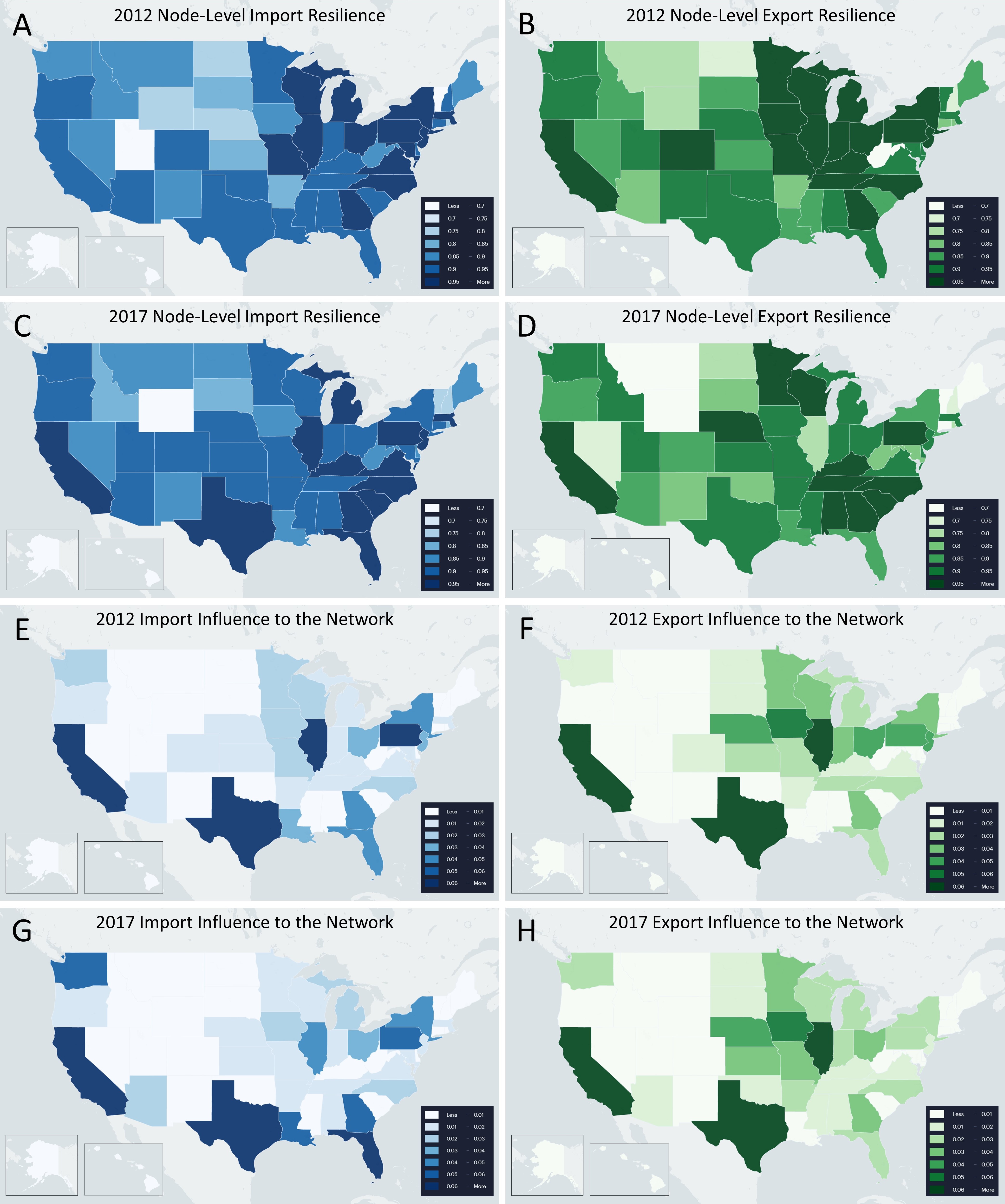}
	\caption{The visualization of the state-level resilience and influence in the agricultural multi-commodity flow network in the U.S. in 2012 and 2017. A-B: node-level import and export resilience of each state in 2012; C-D: node-level import and export resilience of each state in 2017; E-F: import and export influence of each state on the network in 2012; C-D: import and export influence of each state on the network in 2017.}
	\label{fig:resilience_maps}
\end{figure*}

\begin{table}[h]
	\centering
	\caption{List of agricultural commodity types in SCTG.}
	\label{tab:freq}
	\begin{tabular}{p{1cm}p{6.5cm}}
		\toprule
		Code&Description\\
		\midrule
		01 & Live Animals and Fish\\
		02 & Cereal Grains (including seed)\\
		03 & Agricultural Products Except for Animal Feed, Cereal Grains, and Forage Products\\
		04 & Animal Feed, Eggs, Honey, and Other Products of Animal Origin\\
		05 & Meat, Poultry, Fish, Seafood, and Their Preparations\\
		06 & Milled Grain Products and Preparations, and Bakery Products\\
		07 & Other Prepared Foodstuffs, Fats and Oils\\
		08 & Alcoholic Beverages and Denatured Alcohol\\
		\bottomrule
	\end{tabular}
	\label{tab:commodity_types}
\end{table}

\begin{figure*}[h]
	\centering
	\includegraphics[width=\linewidth]{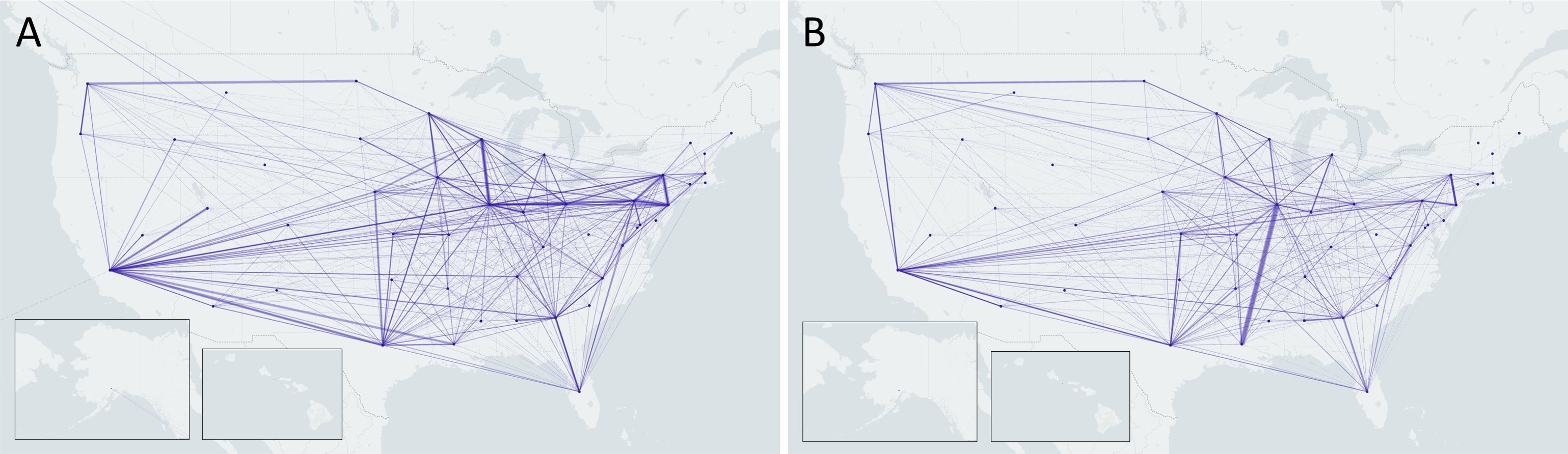}
	\caption{The state-level agricultural multi-commodity flow network in the U.S. in 2012 (left) and 2017 (right).}
	\label{fig:flow_maps}
\end{figure*}

In our experiments, we utilize the CFS-GeoKG and designed metrics to measure both node-level resilience and network-level resilience of the US Multi-Commodity Flow Network. The CFS is conducted every five years by the Bureau of Transportation Statistics (BTS) and the U.S. Census Bureau. CFS provides comprehensive data of domestic freight shipments including commodity type, value, weight, distance shipped, origin and destination, etc. from national-level to state-level. As an example, we focus on the agricultural multi-commodity flows in 2012 and 2017 at differential geographical scales by extracting the data with SCTG code from 01 to 08 (details are listed in Table \ref{tab:commodity_types}) and under different geographical units (i.e., state, division, and region). Following the SCTG Commodity Code List\footnote{https://bhs.econ.census.gov/bhsphpext/brdsearch/scs\_code.html}, we further aggregate code 01-05 (i.e., Agriculture Products and Fish, denoted as $A$) and 06-08 (i.e., Grains and Alcohol, denoted as $B$) to preserve the hierarchical commodity type structure. After loading CFS data into CFS-GeoKG, there are 64 geographical unit entities, including 51 states, 9 divisions, and 4 regions. There are also 3,381 commodity flow entities covering different geographical units, years, and commodity types. The state-level agricultural commodity flows in 2012 and 2017 are visualized on the map in Figure \ref{fig:flow_maps}, respectively. The flow width reflects the total commodity value of each flow.

\begin{figure*}[h]
	\centering
	\includegraphics[width=0.82\linewidth]{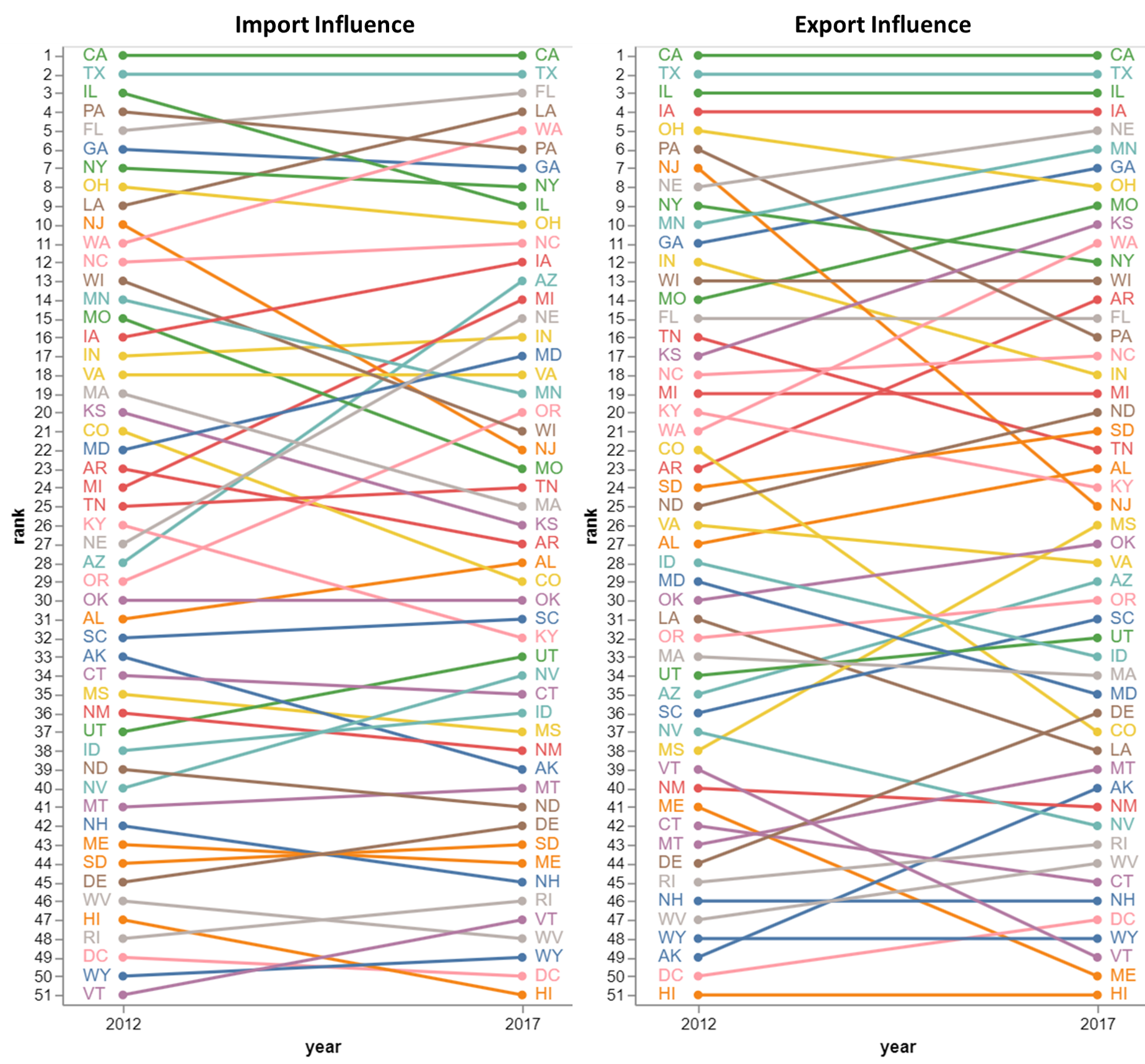}
	\caption{The ranking changes of import influence (left) and export influence (right) of states to the multi-commodity flow network between 2012 and 2017.}
	\label{fig:ranking_plot}
\end{figure*}

We first calculate the node-level import resilience $R^{in}_i$ and export resilience $R^{out}_i$ for each node $i$ at the state level in the agricultural multi-commodity flow network and list the top-10 resilient states in 2017 in Table \ref{tab:node_resilience_2017}. Then we calculate both import influence $I^{in}_i$ and export influence $I^{out}_i$ to the network for each node $i$ and list the top-10 states in 2017 in Table \ref{tab:network_resilience_2017}. The ranking in Table \ref{tab:node_resilience_2017} differs from Table \ref{tab:network_resilience_2017} as node-level resilience mainly reflects the resilience of the nodes themselves to single-source dependence, transport distance, and geographic adjacency, which does not necessarily rely on their total commodity values, while the influence on the network relies heavily on the total commodity value of the nodes. In Figure \ref{fig:resilience_maps}, we further visualize all the node-level import and export resilience for each state in 2012 and 2017 as well as the import and export influence of each state to the entire food flow spatial network in 2012 and 2017. Darker color represents higher resilience or influence. We can observe from the figure that the east-coast states, west-cost states, and mid-west states have higher node-level resilience than other areas. This might be because they have larger and wider supply chains for purchasing and selling agricultural products, making them less dependent on single-sourcing suppliers/customers or geographic constraints, and thereby more resilient to disruption risks. Moreover, we can observe that many top agricultural-producing states such as California, Texas, and mid-west states have high import and export influence on the network, indicating that they contribute high agricultural commodity values, practice robust supply chain strategies, and play important roles in the US food system. In Figure \ref{fig:ranking_plot}, we also show how the ranking of each state's import or export influence on the entire network changes from 2012 to 2017. It shows that the rankings of both the most influential and less influential states in the network do not change a lot, and most of the states in the middle may change rankings within a certain range, which also matches our intuition.

\begin{table}[h]
	\centering
	\caption{The Top-10 states with highest node-level import or export resilience in the network in 2017.}
	\label{tab:freq}
	\begin{tabular}{p{2cm}p{1cm}p{2cm}p{1cm}}
		\toprule
		State & $R^{in}_i$ & State & $R^{out}_i$\\
		\midrule
		North Carolina & 0.968 & California & 0.970 \\
		Kentucky & 0.965 & Georgia & 0.963 \\
		Georgia & 0.965 & Nebraska & 0.962 \\
		Pennsylvania & 0.965 & Kentucky & 0.959 \\
		California & 0.964 & North Carolina & 0.959 \\
		Michigan & 0.963 & Pennsylvania & 0.959 \\
		Florida & 0.961 & South Carolina & 0.957 \\
		Illinois & 0.959 & Alabama & 0.956 \\
		Texas & 0.956 & Minnesota & 0.955 \\
		Virginia & 0.954 & Tennessee & 0.953 \\
		\bottomrule
	\end{tabular}
	\label{tab:node_resilience_2017}
\end{table}

\begin{table}[h]
	\centering
	\caption{The Top-10 most influential states to the import or export of the network in 2017.}
	\label{tab:freq}
	\begin{tabular}{p{2cm}p{1cm}p{2cm}p{1cm}}
		\toprule
		State & $I^{in}_i$ & State & $I^{out}_i$\\
		\midrule
		California & 0.108 & California & 0.132 \\
		Texas & 0.104 & Texas & 0.096 \\
		Florida & 0.062 & Illinois & 0.061 \\
		Louisiana & 0.057 & Iowa & 0.058 \\
		Washington & 0.057 & Nebraska & 0.045 \\
		Pennsylvania & 0.053 & Minnesota & 0.040 \\
		Georgia & 0.052 & Georgia & 0.039 \\
		New York & 0.048 & Ohio & 0.038 \\
		Illinois & 0.046 & Missouri & 0.032 \\
		Ohio & 0.034 & Kansas & 0.031 \\
		\bottomrule
	\end{tabular}
	\label{tab:network_resilience_2017}
\end{table}

We then measure the network-level resilience in 2012 and 2017. Here we report the overall network-level resilience $R_{net}=(R^{in}_{net}+R^{out}_{net})/2$ at three different geographic scales in Table \ref{tab:overall_network_resilience}. The results show that the overall resilience of the agricultural commodity flow network at the state level increased from $0.867$ in 2012 to $0.880$ in 2017 by $1.5\%$, indicating the success in creating a more resilient interstate supply chain in agriculture. At the division level and the region level, however, the overall resilience in 2017 drops a bit (by 0.4\% and 6.7\%) from 2012. This might be related to the increasing concentration of agricultural resources in some specific areas. For example, the Pacific Division, which is the division with the most agricultural influence based on CFS data, continued to increase its import and export influence by 11.6\% and 8.6\% in 2017 compared to 2012, respectively, which leads to a drop in network-level resilience in 2017. This observation suggests that we may need to pay attention to the increasing concentration of agricultural resources on a larger geographic scale and take corresponding measures to enhance the resilience of the overall agricultural supply chain network \citep{karakoc2022food}.

\begin{table}[h]
	\centering
	\caption{The overall network-level resilience in 2012 and 2017 at different geographical scales.}
	\label{tab:freq}
	\begin{tabular}{p{2cm}p{1.5cm}p{1.5cm}p{2cm}}
		\toprule
		Scale & $R^{2012}_{net}$ & $R^{2017}_{net}$ & Changes (\%)\\
		\midrule
		State & 0.867 & 0.880 & +1.5\% \\
		Division & 0.766 & 0.763 & -0.4\% \\
		Region & 0.647 & 0.604 & -6.7\% \\
		\bottomrule
	\end{tabular}
	\label{tab:overall_network_resilience}
\end{table}

\section{discussion}

We acknowledge several limits associated with the current data source, each of which prompts further research directions. First, utilizing CFS data requires that we represent both food imported internationally as well as that sourced within the U.S. The easily identifiable CFS state-level unit of analysis requires further unpacking to reveal geographically relevant foodsheds. This is to say that corn and wheat produced in the Great Plains cover parts of several states with broadly similar production characteristics and relatively few transportation modes to reach processing centers and markets. Likewise, substate regions, such as cranberry production in Wisconsin or Maine, or the vast variety of products (and associated ecological problems) of California’s central valley demand resilience analysis at an appropriate geographic scale.

Second, regions are differentially diverse in their agricultural products and processing activities. This suggests the need for an index of product diversity, benchmarked on other USDA measures, such as the typical consumer market basket or NIH food pyramids to deepen our understanding of regional resilience. Clearly, such measures should account for locally available processing to transform commodities into consumable products.

Third, we need to further address the notion of value. Ease of analysis compels us to reduce the value to a dollar amount associated with the CFS. This reductionist approach allows for ease of analysis and the ability to advance an initial understanding of resilience. A deeper understanding would follow by developing indices of value that relate to nutrition, energy use, or multifunctionality of specific crops. 

Fourth, our measure of resilience is robust given our operationalization of data and objectives, but, again, a deeper understanding of resilience will include a variety of social factors. The notion of resilience can be operationalized at various scales – individuals, households, communities, ecoregions, etc.. A regional operationalization of resilience would embrace multiple public and private sector organizations and the demographically diverse population associated with a region or state. Characteristics of resilience could include communicative connectivity, flexibility, and fungibility of organizational units and subunits such that resilience is represented as a socio-technical characteristic of a region \cite{woods2019essentials, morales2021public}. Another promising approach to resilience is to examine the relative resilience of the distribution infrastructure associated with geography, and further, we could specify the relative resilience of particular subpopulations, such as those in prisons, schools, hospitals, or non-consumer food systems.

Lastly, our point in this paper is to introduce a GeoKG-based approach to measuring resilience and demonstrate its use. Thus, our results, while real and meaningful, are mainly for reference since the parameters we use have not been thoroughly investigated or calibrated. More appropriate parameter settings will alter the resilience measurement results. So, one more task would be to refine our parameter settings and metric designs by exploring the aforementioned research directions.

\section{Conclusion}

In this paper, we introduce a GeoKG-based approach to measuring the resilience of a multi-commodity flow network. The case study shows that CFS-GeoKG well supports the storage and query of geospatial semantics of a multi-commodity flow network, thereby supporting the comprehensive measurement of single-sourcing dependence and the dependence of geographically distant or non-adjacent suppliers/customers. The CFS-GeoKG well supports measuring both node-level and network-level resilience and also helps discover the increasing concentration of agricultural resources at higher geographical levels, implying broad application prospects of GeoKG in multi-commodity supply chain networks.

Our future works include linking CFS-GeoKG to more existing semantic ontologies for better scalability, understanding geographically relevant foodsheds, product diversity index and their relationships with resilience, integrating more promising data and factors into resilience metrics such as social economics, energy use, and nutrition profiles, and investigating the trade-off between resilience and efficiency of multi-commodity flow networks to discover better food supply chain network structures.

\begin{acks}
We acknowledge the funding support from the National Science Foundation funded AI institute [Grant No. 2112606] for Intelligent Cyberinfrastructure with Computational Learning in the Environment (ICICLE). Any opinions, findings, and conclusions or recommendations expressed in this material are those of the author(s) and do not necessarily reflect the views of the funder(s). We also thank Matthew Lange from IC-FOODS for his valuable suggestions.
\end{acks}

\bibliographystyle{ACM-Reference-Format}
\bibliography{references}

\end{document}